\documentstyle[seceq,twocolumn]{jpsj}
\def\sbox#1{\mbox{\scriptsize #1}}

\def\v#1{\mib #1}
\def\simleq{\mbox{\raisebox{-1.0ex}{$\stackrel{<}{\sim}$}}}
\def\simgeq{\mbox{\raisebox{-1.0ex}{$\stackrel{>}{\sim}$}}}

\def\N{\sbox{N}}
\def\C{\sbox{C}}
\def\c{\sbox{c}}

\def\runtitle{Dimer Expansion Study of the Bilayer $J_1$-$J_2$ Model
}

\def\runauthor{Kazuo {\sc Hida}}

\title
{
Dimer Expansion Study of the Bilayer Square Lattice \\ Frustrated Quantum Heisenberg Antiferromagnet
}

\author
{Kazuo {\sc Hida}
\footnote{e-mail: hida@riron.ged.saitama-u.ac.jp}}

\inst
{
Department of Physics, Faculty of Science,\\ Saitama University, Urawa, Saitama 338
}

\recdate
{
\today
}

\abst
{
The ground state of the square lattice bilayer quantum antiferromagnet with nearest ($J_1$) and next-nearest ($J_2$) neighbour intralayer interaction is studied by means of the dimer expansion method up to the 6-th order in the interlayer exchange coupling $J_3$. The phase boundary between the spin-gap phase and the magnetically ordered phase is determined from the poles of the biased Pad\'e approximants for the susceptibility and the inverse energy gap assuming the universality class of the 3-dimensional classical Heisenberg model.  For weak frustration, the critical interlayer coupling decreases linearly with $\alpha (= J_2/J_1)$. The spin-gap phase persists down to $J_3=0$ (single layer limit) for $0.45 \simleq \alpha \simleq 0.65$. The crossover of the short range order within the disordered phase is also discussed.}

\kword
{
bilayer Heisenberg antiferromagnet, frustration, Pad\'e approximant, dimer expansion method, spin-gap state
}

\begin{document}
\sloppy
\maketitle


\section{Introduction}

The spin-1/2 square lattice Heisenberg model is now widely believed to have an antiferromagnetic long range order in the ground state.\cite{em1,chn1,sgh1,gsh1} It is, however, expected that the strong quantum fluctuation in this system may lead to the destruction of the long range order with the help of some additional mechanism. 
In this context, the square lattice antiferromagnetic Heisenberg model with nearest and next-nearest exchange interaction (hereafter called $J_1$-$J_2$ model)\cite{cd1,sachdev1,sgh1,gsh1,ns1,ok1,mpb1,sdt1,nh1,sz1,szp1,ej1,efj1,zhim1,es1,tmgc1} and the bilayer Heisenberg model\cite{mh1,kh1,kh2,mm1,ss1,weih1,gel2,gwho1,mgc1,mny1} have been studied extensively.  Considering the difference of the nature of the mechanism leading to the spin-gap phase in these two models, it must be most interesting to study their interplay in the bilayer $J_1$-$J_2$ model\cite{khfr,dot}. 
 
In the bilayer model, if the interlayer antiferromagnetic coupling is strong enough, the spins on both layers form interlayer singlet pairs and the quantum fluctuation is enhanced leading to the quantum disordered state.  The dimer expansion study of this model has been quite successful\cite{kh2,weih1,gel2,gwho1,mgc1} and it is shown that the transition between the N\'eel phase and the spin-gap phase belongs to the universality class of 3-dimensional classical Heisenberg model. This result is also confirmed by quantum Monte Carlo simulation.\cite{ss1}

On the other hand, in the $J_1$-$J_2$ model, the competition between the nearest neighbour interaction $J_1$ and the nearest neighbour interaction $J_2$ introduces the frustration in the spin configuration which enhances the quantum fluctuation. The conclusion about the presence of the quantum disordered state in this model is, however, still controversial even in the most frustrated regime. 

 In order to apply the dimer expansion method to the single layer $J_1$-$J_2$ model, it is inevitable to start with the dimer configurations which break the translational symmetry as an unperturbed ground state.\cite{sgh1,gsh1} In the bilayer $J_1$-$J_2$ model, the unperturbed ground state can be taken as the interlayer dimers and the translational symmetry of the original Hamiltonian is preserved throughout the calculation. Therefore the bilayer model is more suitable for the dimer expansion study than the single layer model. It is also possible to get insight into the phase transitions in the single layer model by investigating of the asymptotic behavior in the limit of vanishing interlayer interaction.

This paper is organized as follows: The bilayer $J_1$-$J_2$ model Hamiltonian is introduced in the next section. In \S 3, the dimer expansion method\cite{sgh1,gsh1,gsh2,gel1} is applied to this model and the phase diagram is determined using the biased Pad\'e analysis. The last section is devoted to summary and discussion.

\section{Bilayer $J_1$-$J_2$ Model}

The Hamiltonian of the bilayer $J_1$-$J_2$ model is given as follows,

\begin{eqnarray}
\label{eq:ham1}
H &=&J_1\sum_{<i,j>_{nn}}(\v{S}^A_{i} \v{S}^A_{j}+\v{S}^B_{i} \v{S}^B_{j}) \nonumber \\
&+& J_2\sum_{<i,j>_{nnn}}(\v{S}^A_{i} \v{S}^A_{j}+\v{S}^B_{i} \v{S}^B_{j})  \nonumber \\
&+& J_3\sum_i \v{S}^A_{i} \v{S}^B_{i},
\end{eqnarray} 
where $\v{S^{A}_i}$ and $\v{S^{B}_i}$ are the spin operators with magnitude $1/2$ on the $i$-th site of the layer $A$ and $B$, respectively. The expression $\displaystyle\sum_{<i,j>_{nn}}$ and $\displaystyle\sum_{<i,j>_{nnn}}$ denote the summation over the intralayer nearest neighbour pairs and next nearest neighbour pairs, respectively. The last term represents the interlayer coupling. All exchange couplings are assumed to be antiferromagnetic. In the following, we denote the ratios $J_2/J_1 = \alpha$ and $J_3/J_1 = \beta$. In the classical limit, the ground state is the N\'eel state or the collinear state according as $\alpha <\alpha_{\c}$ or $\alpha > \alpha_{\c}$ where $\alpha_{\c}=0.5$.\cite{cd1,khfr}  

\section{Dimer Expansion Method}

In the absence of the intralayer coupling, the ground state is the assembly of independent interlayer dimers. Treating the intralayer coupling
\begin{eqnarray}
\label{eq:ham2}
H_{\sbox{intra}} &=&J_1\left[\sum_{<i,j>_{nn}}\left\{\v{S}_{i}^A \v{S}_{j}^A+\v{S}_{i}^B \v{S}_{j}^B\right\}\right.  \nonumber \\
&+&\left. \alpha\sum_{<i,j>_{nnn}}\left\{\v{S}_{i}^A \v{S}_{j}^A+\v{S}_{i}^B \v{S}_{j}^B\right\} \right],
\end{eqnarray}
as a perturbation, we apply the expansion with respect to $z = \beta^{-1}$ using the method of Gelfand, Singh and Huse\cite{sgh1,gsh1,gsh2} and Gelfand\cite{gel1} for various values of $\alpha$. In order to calculate the staggered susceptibility $\chi_{\N}$ and collinear susceptibility $\chi_{\C}$, we also add the following magnetic field terms with wave number $\v{Q}=(\pi,\pi)$ and $(\pi,0)$, respectively.

\begin{equation}
\label{eq:hamn}
H_{\v{Q}} =\sum_{i}^{N}h\left[S_{i}^{zA}-S_{i}^{zB}\right](-1)^{\v{Q}\v{r}_i},
\end{equation}
and calculate the ground state energy  $ E(h)$ up to the second order in $h$. Here $\v{r}_i$ is the position of the $i$-th site and $N$ is the number of the lattice sites in a layer. The susceptibility is given by,
\begin{equation}
\chi = - \left.\frac{\partial^2 E(h)}{\partial h^2}\right|_{h \rightarrow 0},
\end{equation}
where $\chi$ stands for $\chi_{\N}$ or  $\chi_{\C}$ according as  $\v{Q}=(\pi,\pi)$ or $(\pi,0)$. Using the method of Gelfand,\cite{gel1} we also calculate the expansion series for the single particle excitation energies $\Delta_{\N}$ and $\Delta_{\C}$ at the wave vector $(\pi, \pi)$ and $(\pi, 0)$, respectively.

These quantities are expanded as a power series in $z$ and $\alpha z$ as

\begin{equation}
\label{ser1}
O = \sum_{p=0}^{\infty}\sum_{q=0}^{\infty}c_{p,q}z^{p}(\alpha z)^{q}=\sum_{k=0}^{\infty}c_k(\alpha)z^{k},
\end{equation}  
and
\begin{equation}
\label{eq:coef}
c_k(\alpha)=\sum_{q=0}^{k}c_{k-q,q}\alpha^{q}.
\end{equation}
Here $O$ stands for $\chi_{\N}$, $\chi_{\C}$, $\Delta_{\N}$  and $\Delta_{\C}$. Actually, the coefficients $c_k(\alpha)$ are calculated up to the 6-th order in $\beta^{-1} \equiv J_1/J_3$ for 7 different values of $\alpha$ and $c_{p,q}$'s are calculated by inverting the relation (\ref{eq:coef}). 

The ratio series of these series are, however, ill-behaved except for the close neighbourhood of $\alpha=0$. In order to locate the phase boundary as precisely as possible from the limited data, we assume that the phase transition of the present model belongs to the universality class of 3-dimensional classical Heisenberg model for which $\chi \sim (\beta-\beta_{\c})^{-\gamma}$ and $\Delta \sim (\beta-\beta_{\c})^{\nu}$ with $\gamma \simeq 1.4$ and $\nu \simeq 0.71$\cite{fh1} even in the presence of frustration. This is expected to be valid because the Berry phase term always cancel between the two layers even if it exists in the single layer model and the remaining long wave length action is given by the 3-dimensional $O(3)$ nonlinear $\sigma$ model,\cite{ej1} Thus we obtain the biased $[L,M]$ Pad\'e approximants for each value of $\alpha$ as,
\begin{equation}
O^{1/\lambda}[L,M] = \frac{\sum_{l=0,L}p^{O}_l z^l}{\sum_{i=0,m}q^{O}_m z^m}, 
\end{equation}
with $q^{O}_0 = 1$ and $L+M \leq 6$ where $\lambda$ stands for $\gamma$ and $-\nu$.  From the poles $z_{\c}$ of the approximants for  $\chi_{\N}$ and $\Delta_{\N}$ ($\chi_{\C}$ and $\Delta_{\C}$), we determine the critical values $\beta_{\c}=1/z_{\c}$ of the phase transition between the N\'eel(collinear) phase and the spin-gap phase for each value of $\alpha$. These approximants behave as
\begin{equation}
O^{1/\lambda}[L,M] \sim \frac{A^{O}_{\c}}{z_{\c} - z}= \frac{B^{O}_{\c}}{\beta-\beta_{\c}},
\end{equation}
in the neighbourhood of poles $z = z_{\c}$.  Depending on $L$ and $M$, we find many poles which are rather scattered. Among them, we only accept the positive poles with smallest $z_{\c}$ (largest $\beta_{\c}$) and positive amplitudes. The poles with amplitudes $A^{O}_{\c}$ less than the cut-off value $\epsilon_A = 0.01 \sim 0.1$ are discarded as spurious. Figures \ref{fig1}(a), (b) and (c) show the $\alpha$-dependence of the poles of the 6-th, 5-th and 4-th order approximants with $L = M, M \pm 1$. 

For small $\alpha$, the critical value of $\beta$ decreases linearly with $\alpha$. This behavior is common for all poles shown in Fig.\ref{fig1} and consistent with other calculations\cite{dot,khfr}. For general values of $\alpha$, it is physically reasonable to assume that the critical value of $\beta$ decreases (increases) with the increase of $\alpha$ for N\'eel(collinear)-spin-gap transition. Some poles, however, show the opposite behavior as shown in Fig. \ref{fig1}. We assume these poles are physically meaningless. If these poles are omitted, the qualitative features of the phase diagram is common for all approximants. Namely, no acceptable real positive poles are found in the interval $0.45 \simleq \alpha \simleq 0.65$ indicating that  the spin-gap phase is stable for the single layer model in this interval of $\alpha$. The N\'eel (collinear) ordered state appears for  $\alpha \simleq 0.45$ ($\alpha \simgeq 0.65$). This is consistent with the exact diagonalization results\cite{sz1,szp1} and some approximate estimations,\cite{sgh1,gsh1,cd1,sachdev1,ej1,efj1,zhim1,es1,tmgc1} although the precise value of the critical $\alpha$ depends on the method used. On the other hand, the corresponding amplitude $B^{\Delta}_{\c}$ does not show any singular behavior as $\beta_{\c} \rightarrow 0$ on these poles as shown in Fig. \ref{fig2} for the $[3,3]$ approximants of $\Delta_{\N}$ and $\Delta_{\C}$. This suggests that the universality class of the transition in the single layer model belongs to the same universality class as the bilayer model. This is consistent with the prediction that the Berry phase term is {\it dangerously irrelevant} even if it exists in the single layer model,\cite{chub1}

Using the [3,3] approximant, the $\beta$-dependence of the energy gap is shown in Fig. \ref{fig3} for various values of $\alpha$. It is clear that the energy gap $\Delta_{\N}$ at $(\pi,\pi)$ increases and  $\Delta_{\C}$ at $(\pi, 0)$ decreases with $\alpha$. The crossover point $\alpha_{\sbox{cr}}(\beta)$, at which  the position of the smallest gap shifts from $(\pi, \pi)$ to $(\pi, 0)$, varies with $\beta$ as shown in Fig. \ref{fig4}. For large values of $\beta$, $\alpha_{\sbox{cr}}$ is close to 0.5 at which the classical ground state changes from the N\'eel state to the collinear state. It shifts to 0.576 as $\beta$ tends to 0. It should be noted that the phase boundary between the N\'eel phase and the collinear phase is also shifted to 0.6 in the modified spin wave approximation\cite{ns1,khfr} for small $\beta$. This can be interpreted in the following way. The dominant short range order is N\'eel type or the collinear type according as $\alpha \simleq 0.576$ or $\simgeq 0.576$ for small $\beta$. In the modified spin wave approximation, the corresponding long range order is established because of the underestimation of the quantum fluctuation.

\section{Summary and Discussion}

The spin-1/2 bilayer $J_1$-$J_2$ model is studied by means of the dimer expansion method and the ground state phase diagram is obtained by the biased Pad\'e analysis assuming the universality class of the 3-dimensional Heisenberg model. For small interlayer coupling, the critical value of $\beta$ for the transition between the N\'eel phase and the spin-gap phase decreases linearly with $\alpha$. Within the available data, the spin-gap phase remains stable down to $\beta = 0$ for $0.45 \simleq \alpha \simleq 0.65$ which is consistent with some of earlier estimations. It is also suggested that the phase transitions in the bilayer model and the single layer model belong to the same universality class.

The excitation gaps at  $(\pi,\pi)$ and  $(\pi,0)$ are calculated as a function of $\alpha$ and $\beta$ using the [3,3] Pad\'e approximant. It is shown that the minimum gap shifts from $(\pi,\pi)$ to  $(\pi,0)$ at $\alpha_{\sbox{cr}}$ which is close to 0.5 for large $\beta$ and grows to 0.576 as $\beta \rightarrow 0$. 

At the first glance, these results appear to contradict with the results of the modified spin wave approximation\cite{ns1,khfr} which predicts the absence of the spin-gap phase in the single layer model. This method also predicts substantially large critical value of $\beta$. These are due to the underestimation of the quantum fluctuation in the modified spin wave approximation. From this point of view, the present results are consistent with the modified spin wave results\cite{khfr} if the long range orders found in the latter approximation is reinterpreted as the corresponding short range orders.

Needless to say, the present conclusion is far from conclusive. The order of the expansion is still too low and only small number of approximants are available. Higher order calculation is required to obtain more reliable results.  Unfortunately, however, the number of the dimer expansion graphs becomes enormous due to the presence of next nearest interaction. For example, it amounts 64303 even for $k=6$ (present calculation) and the CPU time consumed for the calculation is nearly 10 hours on FACOM VPP500 supercomputer for each $\alpha$. For $k=7$ the number of the graphs increase by more than a factor of 10 and the calculation of each graph requires even more computational time.

The numerical simulation is performed using the FACOM VPP500 at the Supercomputer Center, Institute for Solid State Physics, University of Tokyo and the HITAC S820/80 at the Information Processing Center of Saitama University.  This work is supported by the Grant-in-Aid for Scientific Research from the Ministry of Education, Science, Sports and Culture.

\newpage
\begin{figure}
\caption{The poles $\beta_{\c}=1/z_{\c}$ for (a) $N=6$,  (b) $N=5$ and $N=4$. The symbols are defined in the figure. The points in the left (right) half of the figures are the poles of  $\chi_{\N}^{1/\gamma}$ and $\Delta_{\N}^{-1/\nu}$ ($\chi_{\C}^{1/\gamma}$ and $\Delta_{\C}^{-1/\nu}$).}
\label{fig1}
\caption{The amplitude $B^{\Delta}_{\c}$ for $[3,3]$ approximants of $\Delta_{\N}$ and  $\Delta_{\C}$. The symbols are common with Fig. 1(a). }
\label{fig2}
\caption{The $\beta$-dependence of the energy gaps $\Delta_{\N}$ and  $\Delta_{\C}$ for various values of $\alpha$ based on the [3,3] Pad\'e approximant. The symbols are defined in the figure. }
\label{fig3}
\caption{The crossover point $\alpha_{\sbox{cr}}$ at which the minimum energy gap shifts from $(\pi,\pi)$ to $(\pi,0)$ based on the [3,3] Pad\'e approximant. }
\label{fig4}
\end{figure}

\begin{thebibliography}{999}
\bibitem{em1}
E. Manousakis: Rev. Mod. Phys.  {\bf 63} (1991) 1 and references therein.
\bibitem{chn1}
S. Chakravarty, B. I. Halperin and D. R. Nelson
: Phys. Rev. Lett.  {\bf 60} (1988) 1057; Phys. Rev.  B{\bf 39} (1989) 2344.
\bibitem{sgh1}
 R. R. P. Singh, M. P. Gelfand and D. A. Huse: Phys. Rev. Lett. {\bf 60}
 (1988) 2484.
\bibitem{gsh1}
M. P. Gelfand, R. R. Singh and D. A. Huse: Phys. Rev. {\bf B40}
 (1989) 10801.
\bibitem{cd1}
P. Chandra and B. Doucot: Phys. Rev. B38 (1988) 9335.
\bibitem{sachdev1}
S. Sachdev and R. N. Bhatt: Phys. Rev. {\bf B41} 9323 (1990).
\bibitem{ns1}
H. Nishimori and Y. Saika: J. Phys. Soc. Jpn. {\bf 59} (1990) 4454.
\bibitem{ok1}
T. Oguchi and H. Kitatani:  J. Phys. Soc. Jpn.  {\bf 59} (1990) 3322.
\bibitem{mpb1}
J.  Mila, D.  Poilblanc and C.  Bruder: Phys.  Rev. B{\bf 43} (1991) 7891.
\bibitem{sdt1}
K. Sano, I.Doi and K. Takano: J. Phys. Soc. Jpn. {\bf 60} (1991) 3807.
\bibitem{nh1} 
T. Nakamura and N. Hatano: J. Phys. Soc. Jpn. {\bf 62} (1993) 3063.

\bibitem{sz1}
H. J. Schulz and T. A. L. Ziman: Europhys. Lett. {\bf 18} (1992) 355. 
\bibitem{szp1}
H. J. Schulz, T. A. L. Ziman, D. Poilblanc: in {\it Magnetic System with Competing Interaction ed. H. T. Diep} World Scientific (1994) 120.
\bibitem{ej1}
T.  Einarsson and H. Johannesson: Phys. Rev. B{\bf 43} (1991) 5867.
\bibitem{efj1}
T.  Einarsson, P.  Fr\"ojdh  and H. Johannesson: Phys. Rev. 
B{\bf 45} (1992) 13121.
\bibitem{zhim1}
M. E. Zhitomirsky and K. Ueda: Phys. Rev. {\bf B54} (1996) 9007.
\bibitem{es1}
T.  Einarsson and H. J. Schulz: Phys. Rev. {\bf B51} 6151 (1995).
\bibitem{tmgc1}
A. E. Trumper, L. O. Manuel, C. J. Gazza and H. A. Ceccatto: Phys. Rev. Lett. {\bf 78} 2216 (1997).
\bibitem{mh1}
T. Matsuda and K. Hida: J. Phys. Soc. Jpn.  {\bf 59} (1990) 2223.
\bibitem{kh1}
K. Hida: J. Phys. Soc. Jpn.  {\bf 59} (1990) 2230.
\bibitem{kh2}
K. Hida: J. Phys. Soc. Jpn.  {\bf 61} (1992) 1013.
\bibitem{mm1}
A. J. Millis and H. Monien: Phys. Rev. Lett. {\bf 70} (1993) 2810; Phys. Rev. {\bf B50} 16606 (1994).
\bibitem{ss1}
A. W. Sandvik and D. J. Scalapino: Phys. Rev. Lett. {\bf 72} (1994) 2777.
\bibitem{weih1}
Z. Weihong: Phys. Rev. {\bf B55} (1997) 12267.
\bibitem{gel2}
 M. P. Gelfand: Phys. Rev. {\bf B53} (1996) 11309.
\bibitem{gwho1}
 M. P. Gelfand, Z. Weihong, C. J. Hamer and J. Oitmaa: cond-mat/9705201.
\bibitem{mgc1}
Y. Matsushita, M. P. Gelfand and C. Ishii: J. Phys. Soc. Jpn. {\bf 66} (1997) 3648.
\bibitem{mny1}
T. Miyazaki, I. Nakamura and D. Yoshioka:  Phys. Rev. {\bf B53} (1996) 12206.

\bibitem{dot}

A. V. Dotsenko: Phys. Rev. B{\bf 52} (1995) 9170.

\bibitem{khfr}
K. Hida: J. Phys. Soc. Jpn.  {\bf 65} (1996) 594.

\bibitem{gsh2}
 M. P. Gelfand, R. R. P. Singh and D. A. Huse: J. Stat. Phys. {\bf 59} (1990) 1093.
\bibitem{gel1}
 M. P. Gelfand: Solid State Comm. {\bf 98} (1996) 11.
\bibitem{fh1}
M. Ferer and A. Hamid-Aidinejad: Phys. Rev. {\bf B34} (1986) 6481
\bibitem{chub1}
A. V. Chubukov, S. Sachdev and J. Ye: Phys. Rev. {\bf B49} (1994) 11919.
\end{thebibliography}
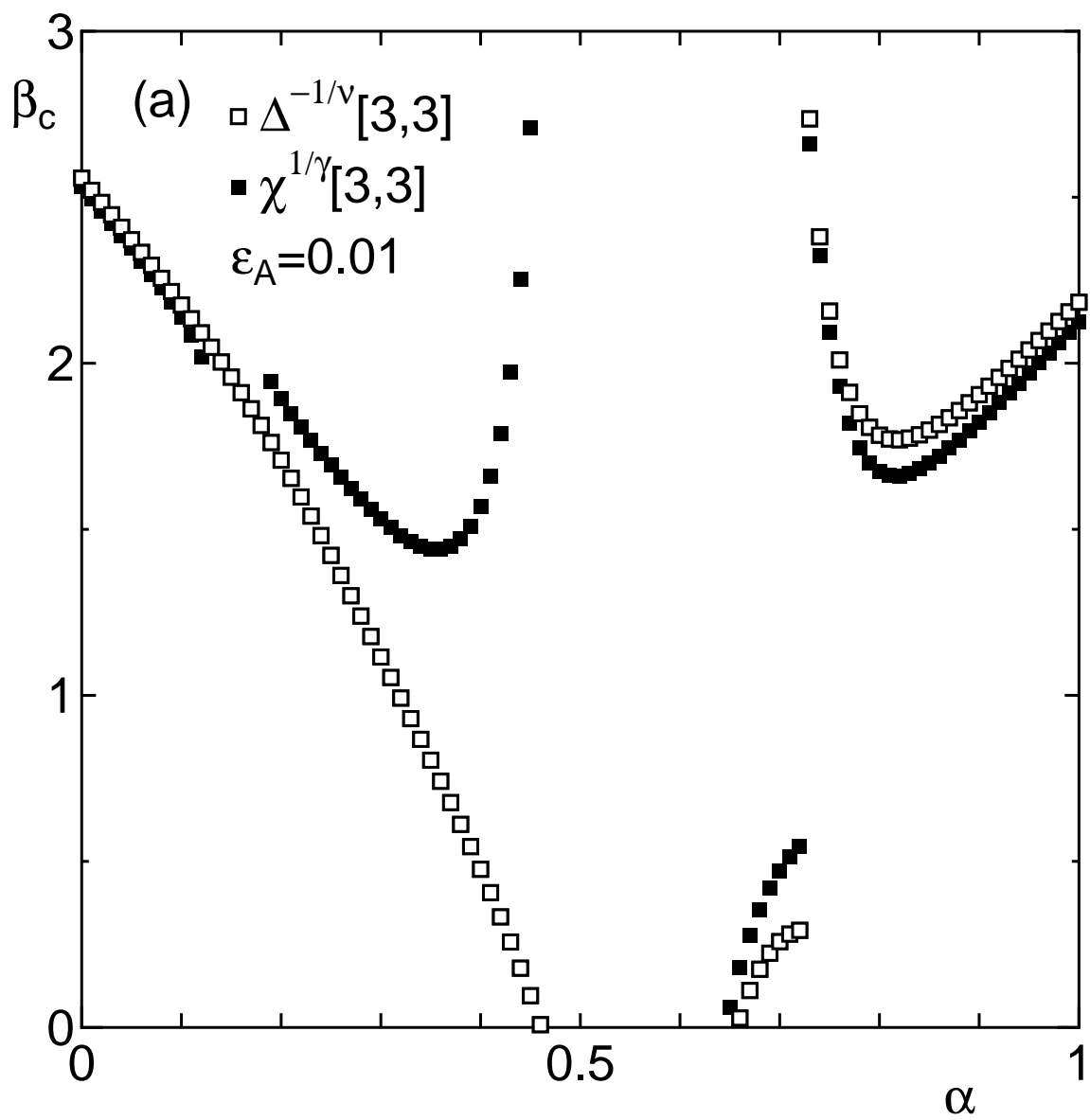
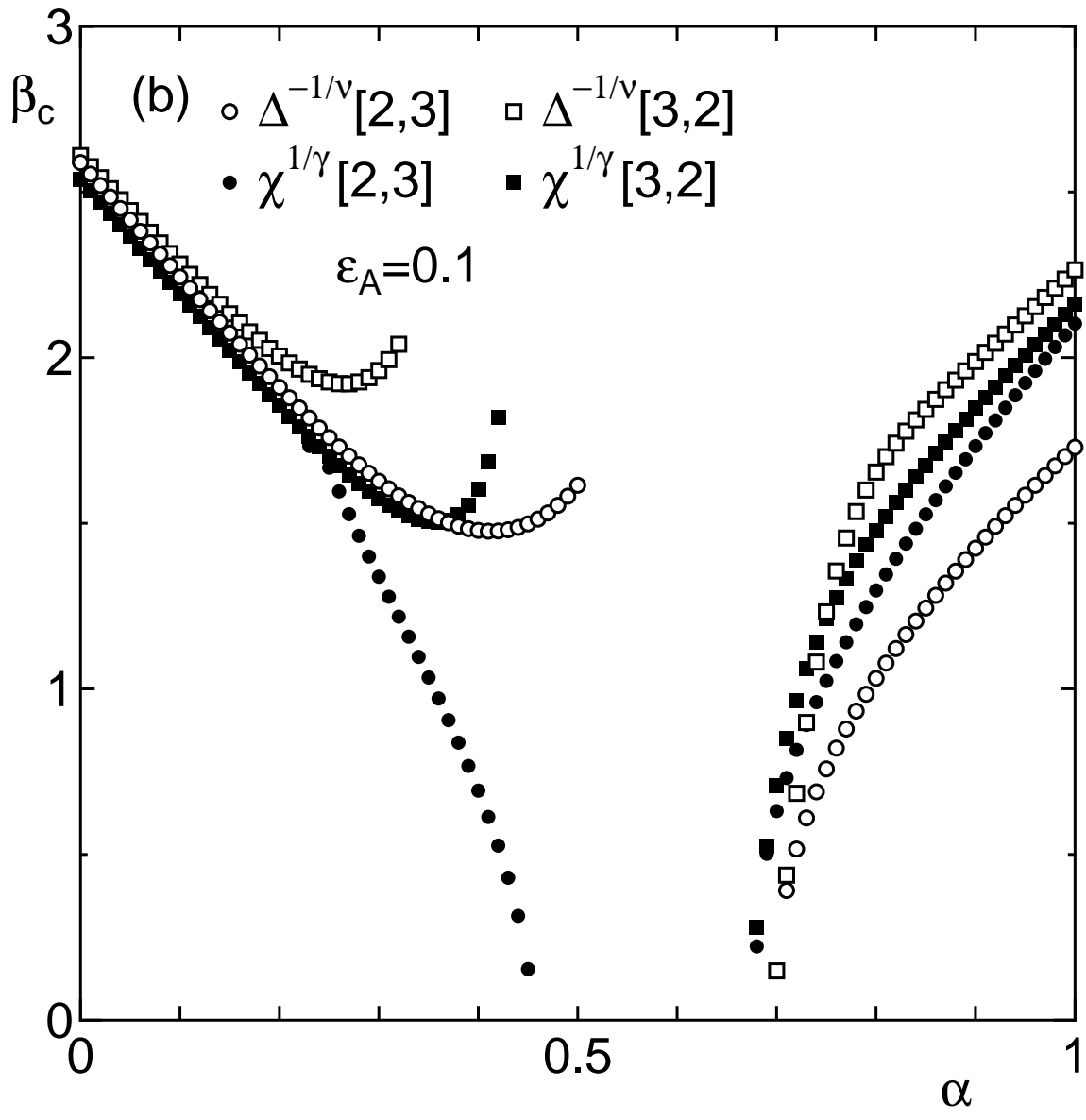
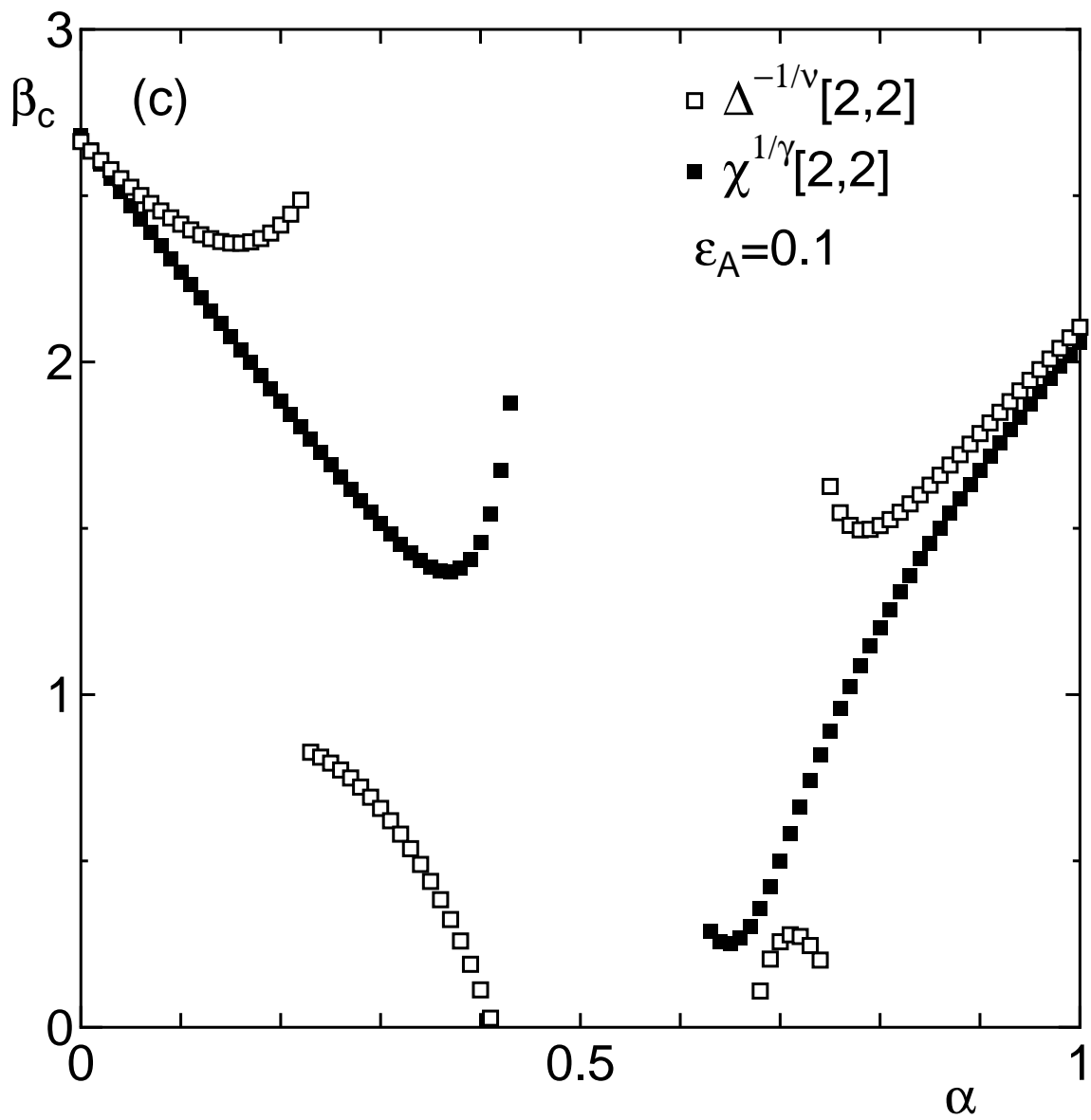
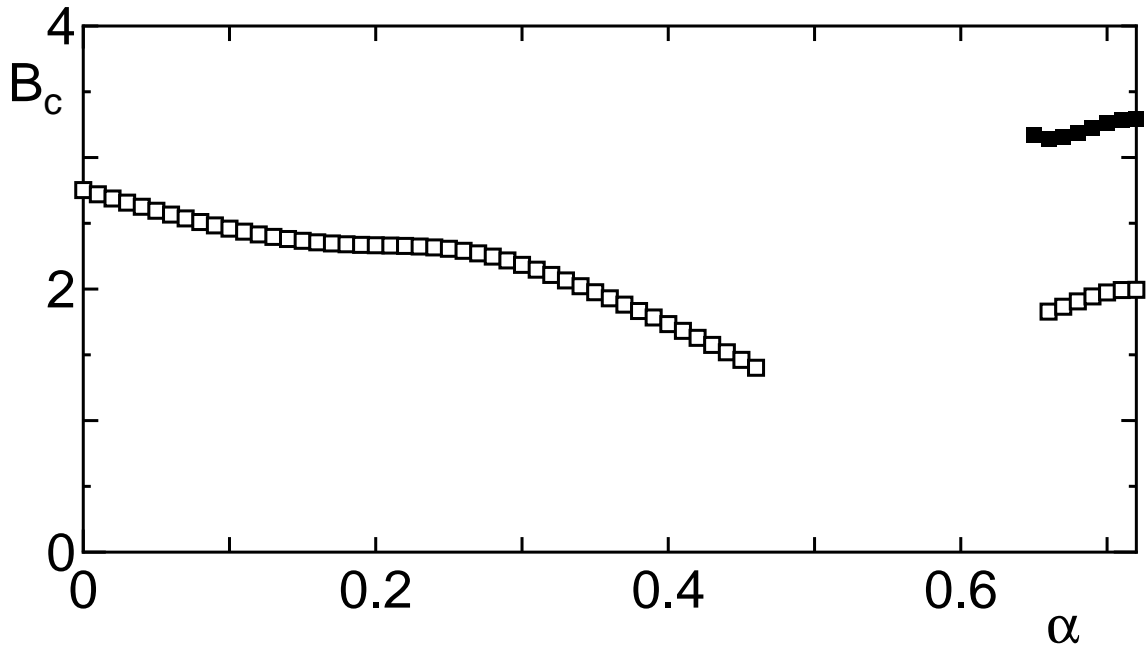
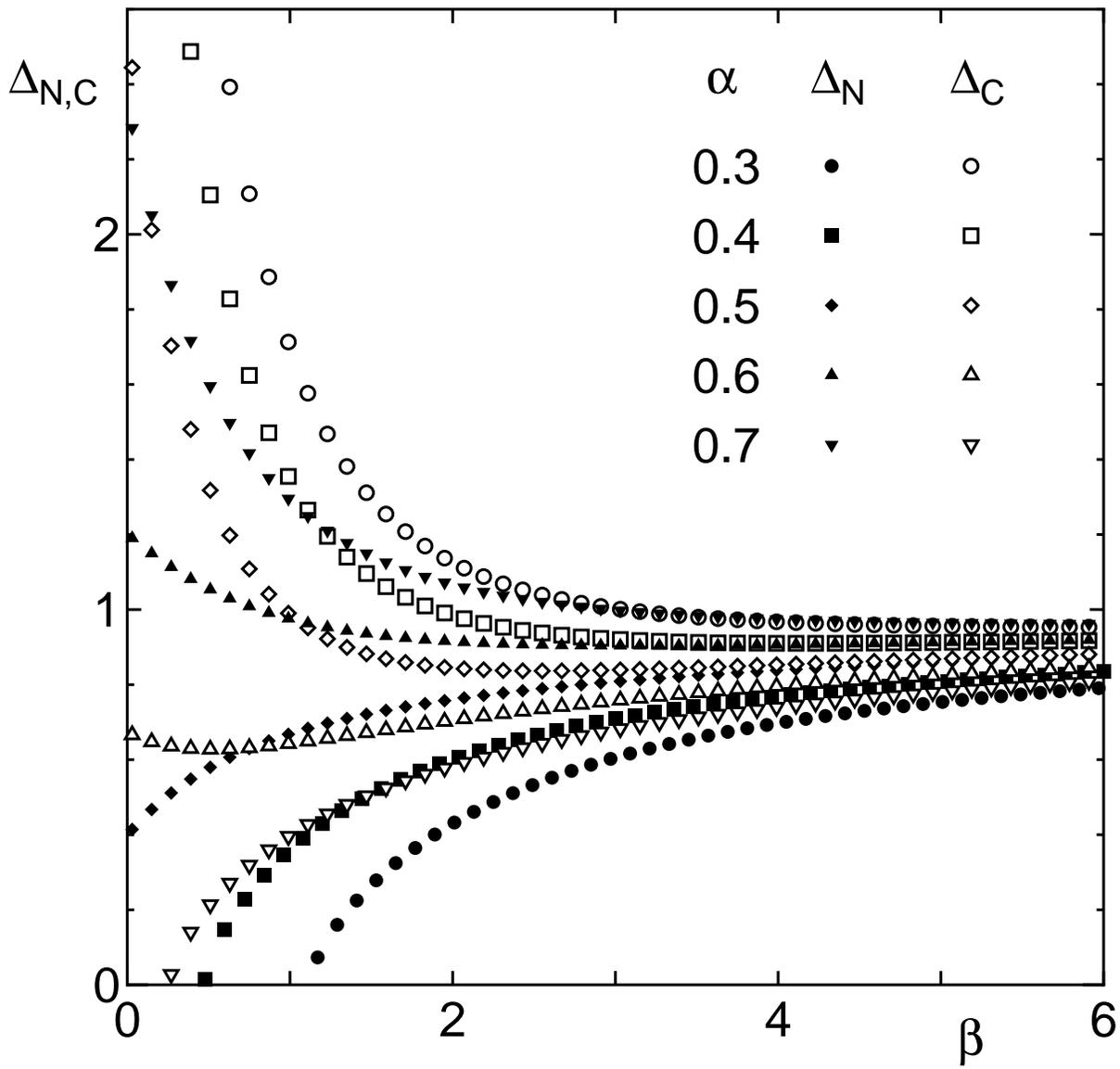
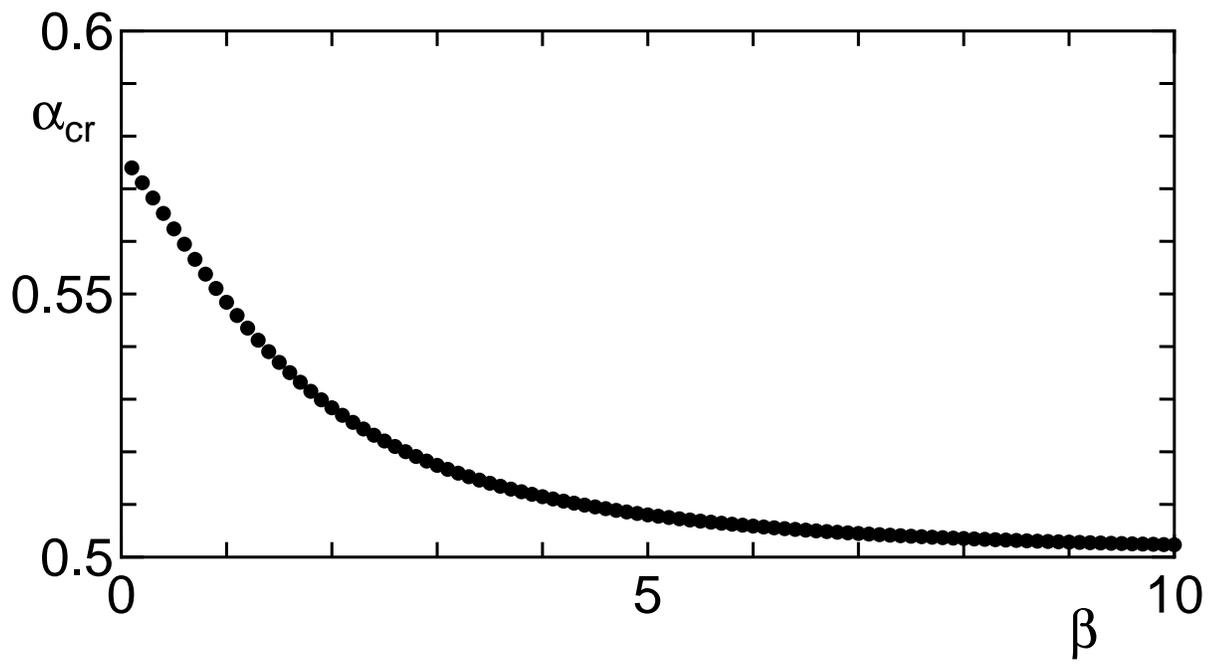
\end{document}